\documentclass[aps,twocolumn,prl,preprintnumbers,amsmath,amssymb,superscriptaddress]{revtex4-1}

\usepackage{graphicx}
\usepackage{bm}
\usepackage{hyperref}
\usepackage{IEEEtrantools}
\usepackage{natbib}
\usepackage{float}

\restylefloat{table}

\begin{document}

\title{Magneto-Terahertz Response and Faraday Rotation from Massive Dirac Fermions in the Topological Crystalline Insulator Pb$_{0.5}$Sn$_{0.5}$Te}

\author{Bing Cheng}
\affiliation{Department of Physics and Astronomy, The Johns Hopkins University, Baltimore, Maryland 21218, USA}

\author{P. Taylor}
\affiliation{Sensors and Electron Devices Directorate, Army Research Laboratory,  2800 Powder Mill Rd, Adelphi, MD 20783 }

\author{P. Folkes}
\affiliation{Sensors and Electron Devices Directorate, Army Research Laboratory,  2800 Powder Mill Rd, Adelphi, MD 20783 }

\author{Charles Rong}
\affiliation{Sensors and Electron Devices Directorate, Army Research Laboratory,  2800 Powder Mill Rd, Adelphi, MD 20783 }

\author{N. P. Armitage}
\affiliation{Department of Physics and Astronomy, The Johns Hopkins University, Baltimore, Maryland 21218, USA}
\date{\today}

\date{\today}

\begin{abstract}

 Pb$_{1-x}$Sn$_x$Te has been shown to be an interesting tunable topological crystalline insulator system.  We present a magneto-terahertz spectroscopic study of thin films of Pb$_{0.5}$Sn$_{0.5}$Te. The complex Faraday rotation angle and optical conductivity in the circular basis are extracted without any additional assumptions.  Our quantitative measures of the THz response allow us to show that the sample studied contains two types of bulk carriers. One is $p$-type and originates in 3D Dirac bands. The other is $n$-type and appears to be from more conventional 3D bands. These two types of carriers display different cyclotron resonance dispersions. Through simulating the cyclotron resonance of hole carriers, we can determine the Fermi energy and Fermi velocity. Furthermore, the scattering rates of $p$-type and $n$-type carriers were found to show opposite field dependences, which can be attributed to their different Landau level broadening behaviors under magnetic field. Our work provides a new way to isolate real topological signatures of bulk states in Dirac and Weyl semimetals.

\end{abstract}

\maketitle

The search for exotic fermionic states in condensed matter is a very active area of physics.  With the advent of graphene \cite{Novoselov04}, topological insulators \cite{Hsieh08,Hasan11} and their topological protected surface states (TSSs), massless Dirac fermions in two dimensions (2D) have been found to be a source of much interesting physics.  Topological band theory has successfully predicted large numbers of topological states in quantum materials. Beyond strict 2D and the 2D surface states on the surface of topological insulators, a number of 3D topological materials, such as Dirac and Weyl semimetals, have been proposed and begun to attract attention \cite{NPA18}. In Dirac semimetals, 3D band crossings with linear dispersion are protected by crystal symmetries and survive on high symmetry lines in momentum space, which makes the low-energy physics near the crossings described by the relativistic Dirac equation \cite{Cd2As3_LDA13}.  Such Dirac fermions possess non-trivial band geometric effects such as large Berry curvature.  When breaking inversion or time reversal symmetry, a two-fold degenerate Dirac node may split into two Weyl nodes with different chiralities \cite{TaAs_LDA15,TaAs_LDA15_2}. A number of exotic phenomena, such as Fermi arc surface states and the chiral anomaly may be exhibited in these topological semimetals.

Among these topological materials, Pb$_{1-x}$Sn$_x$Te is particularly interesting. PbTe is a conventional thermoelectric semiconductor. By carefully choosing {\em x}, the bands around four L points can be inverted \cite{TCI_LDA12}. Given that there are four band inversions in the Brillouin zone, Pb$_{1-x}$Sn$_x$Te is Z$_{2}$ trivial.  However, by taking mirror symmetries into consideration, a new integer topological invariant \textit{mirror} Chern number can be defined.  In Pb$_{1-x}$Sn$_x$Te for $x \gtrapprox 0.3 $, this invariant is non-zero making it a topological crystalline insulator \cite{TCI_theory11}.  This gives an even number of TSSs in the surface Brillouin zone, which are protected by this crystalline symmetry. Besides novel surface states, recent transport experiments have also shown that the bulk states in Pb$_{1-x}$Sn$_x$Te are interesting. The observation of the non-degenerate first Landau level bulk states by thermoelectric measurements demonstrates that the low-energy physics of the bulk can be well described by 3D massive Dirac fermions \cite{TCI_Tian13}.  With applying moderate pressure, dc transport measurements show Pb$_{1-x}$Sn$_x$Te can be driven into a unusual bulk metallic phase that is consistent with a Weyl semimetal\cite{TCI_Tian15}.   All these findings make Pb$_{1-x}$Sn$_x$Te notable as it can be regarded as a parent compound to different kinds of relativistic fermions: 2D massless Dirac fermions, 3D massive Dirac fermions and 3D Weyl fermions.

In this work, we use time-domain terahertz spectroscopy (TDTS) to study several Pb$_{0.5}$Sn$_{0.5}$Te thin films with different thicknesses. We observe coherent low-energy transport and large magneto-optical responses that we show are signatures of massive bulk Dirac fermions. Especially in the thickest sample (325 nm), bulk states are dominant.  We found, besides the $p$ type carriers, $n$-type carriers are also present. These two types of charges show distinct cyclotron resonances and field dependences of the scattering rates, which can be attributed to the different non-trivial properties of their band structures.


The real part of the zero field optical conductance $G_{1}$ of three Pb$_{0.5}$Sn$_{0.5}$Te thin films is displayed in Fig. 1(c). At 6 K, $G_{1}$ is Drude-like for all samples, but shows strong thickness dependence. $G_{1}$ of the 325 nm film is nearly ten times bigger than $G_{1}$ of the 50 nm film. In this TCI system, both surface and bulk states may coexist because the Fermi level is in the bulk states.  This makes it is difficult to distinguish bulk and surface through transport experiments. However, in the thin film with the thickness of 325 nm, bulk states are obviously dominant and the optical spectral weights of surface states are estimated to be less than 5\% [See supplementary material]. In what follows we will concentrate on the thickest sample to study the interesting bulk states and leave the investigation of surface states in TCIs to future work. 

Fig. 1(d) shows the real part of the optical conductivity $\sigma_{1}$ for the 325 nm thick film at different temperatures. At 6 K, $\sigma_{1}$ shows a narrow Drude peak. Because of the lack of obvious phonon features in the measured frequency range, we can use the pure Drude expression $ \omega_p^2 \tau  /1- i \omega \tau$ with a temperature dependent scattering rate (1/2$\pi$$\tau$) and plasma frequency ($\omega_p^2$=$ne^{2}$/$m^{*}$) to fit $\sigma$. One can see the real and imaginary parts of optical conductivity [Fig. 1(e)] at 6 K can be well reproduced by a single Drude oscillator. The plasma frequency $\omega_p/2\pi$ and scattering rate 1/2$\pi$$\tau$ are found to be 125 THz and 1.3 THz, respectively. As temperature increases, the plasma frequency shows little temperature dependence, but the scattering rate is an increasing function of temperature, consistent with a typical metallic behavior.

Magneto-terahertz spectroscopic data were taken in Faraday geometry (FG) [See Fig. 1(a)]. In FG with finite magnetic field, we must convert the linear-basis transmission matrix to the eigenpolarization circular-basis to get conductance as discussed in the SM and in Ref. \cite{armitage2014constraints}.  It is quite illustrative to display the response to right- (R) and left-hand (L) polarized light as positive and negative frequencies respectively and hence the conductivity in the circular basis is single continuous function which can be smoothly extended through zero frequency as shown in Fig. 2(a).   This follows from the fact that we may understand R and L polarized light as having time dependencies that go as $e^{ \mp i \omega t}$ respectively. At zero field, $\sigma_{cir1} $ is a function peaked at zero frequency. With increasing positive field,  the resonance smoothly moves to finite positive frequency and the conductivity is suppressed on the negative frequency side.   This shift of the resonance can be identified as a cyclotron resonance (CR) of $p$-type free carriers. 

\begin{figure}[t]
\includegraphics[clip,width=3.5in]{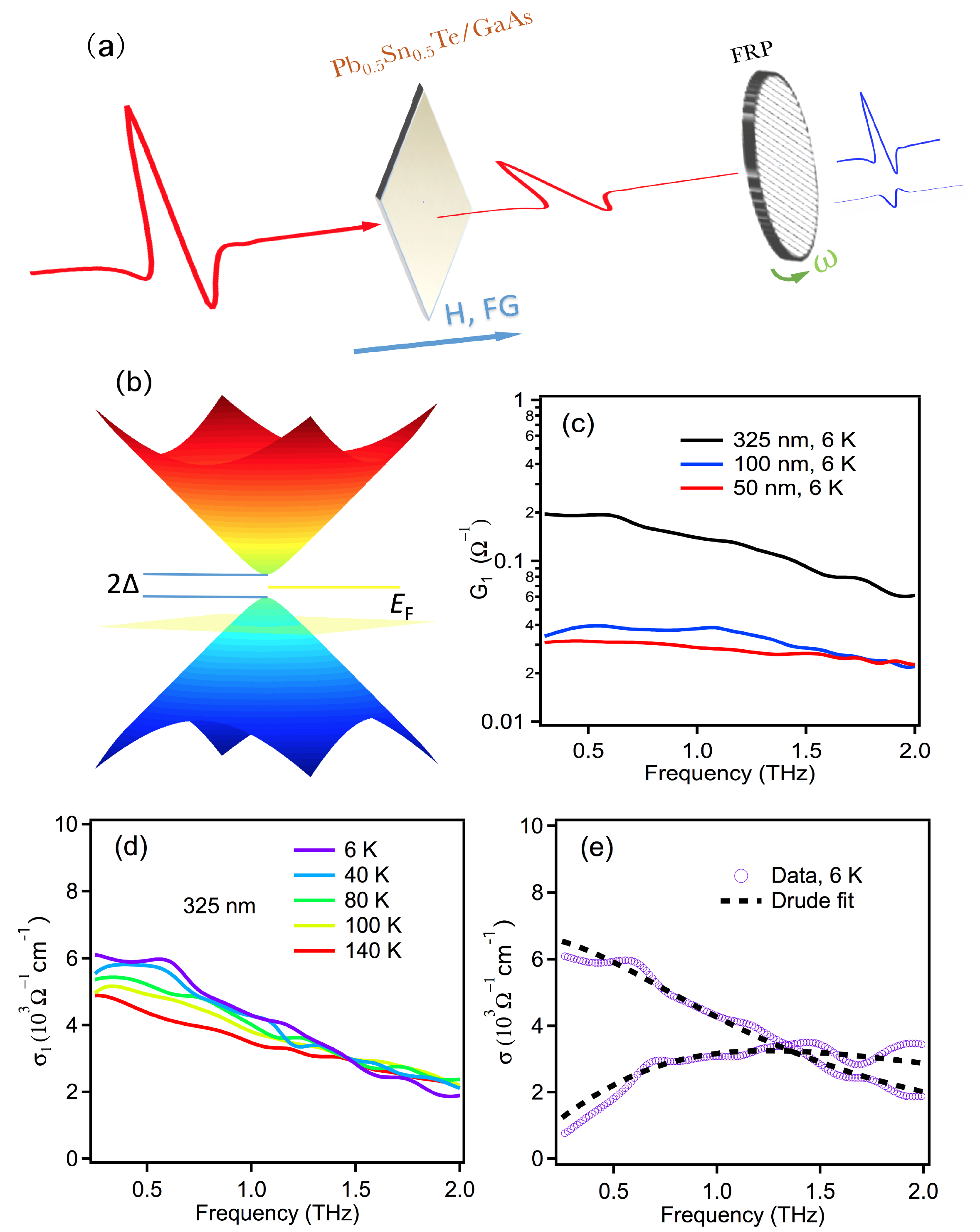}
\caption{(Color online) (a) Schematic of experimental setups used to collect data in Faraday geometry.    (b) Schematic of 3D massive Dirac bands. (c) Real part of optical conductances of thin films with three different thicknesses at 6 K. (d) Temperature-dependent real part of optical conductivity of thin film with thickness of 325 nm. (e) Drude model fit of complex optical conductivity at 6 K. }
\label{xxx}
\end{figure}

\begin{figure}[t]
\includegraphics[clip,width=3.4in]{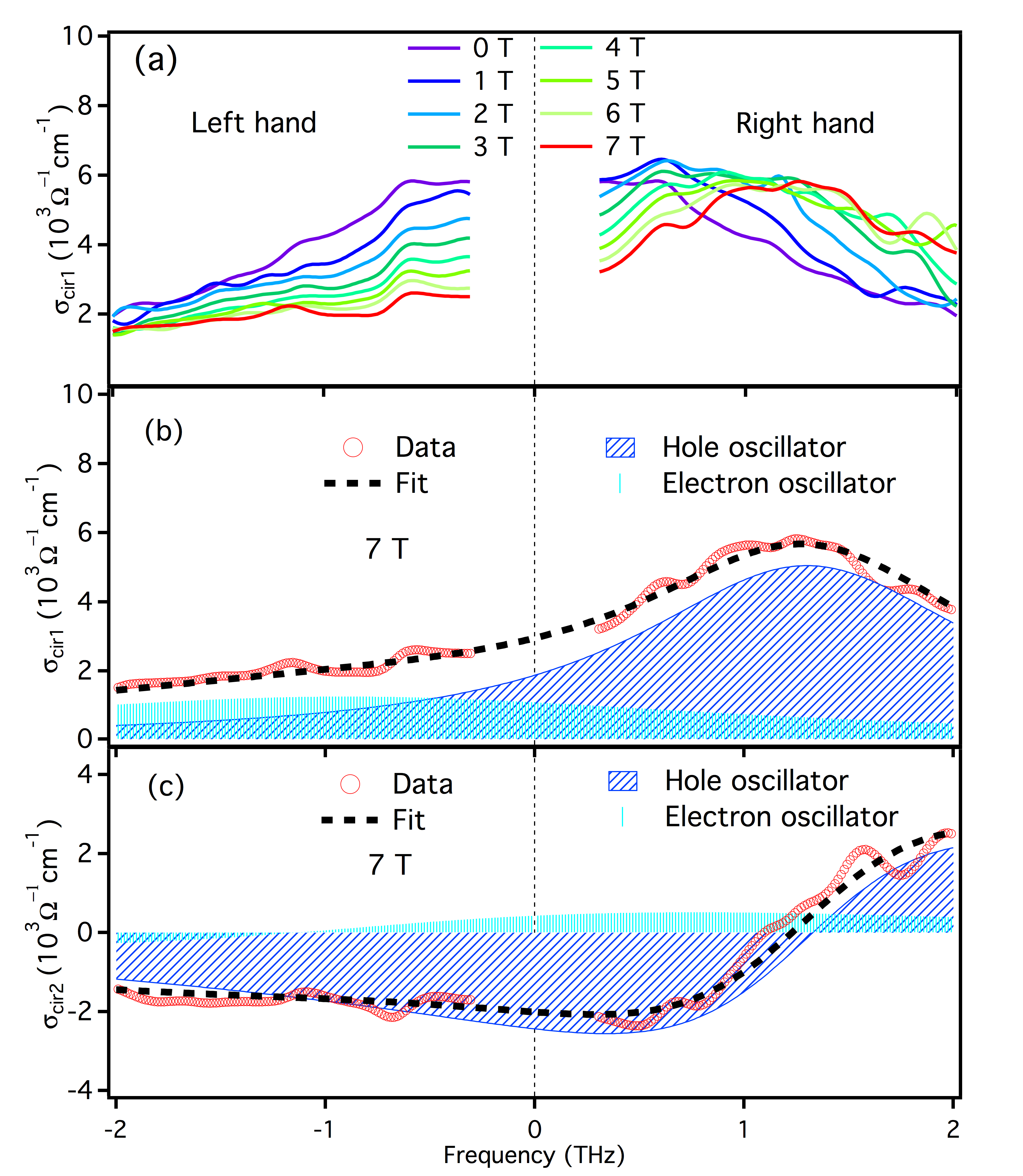}
\caption{(Color online) (a) Real part of the magneto-optical conductivity under circular basis at 6 K. Right-hand and left-hand optical conductivity are displayed as positive and negative frequencies respectively. (b) and (c) Drude fits of real and imaginary parts of optical conductivity under circular basis at 7 T.  }
\end{figure}

\begin{figure}[t]
\includegraphics[clip,width=3.5in]{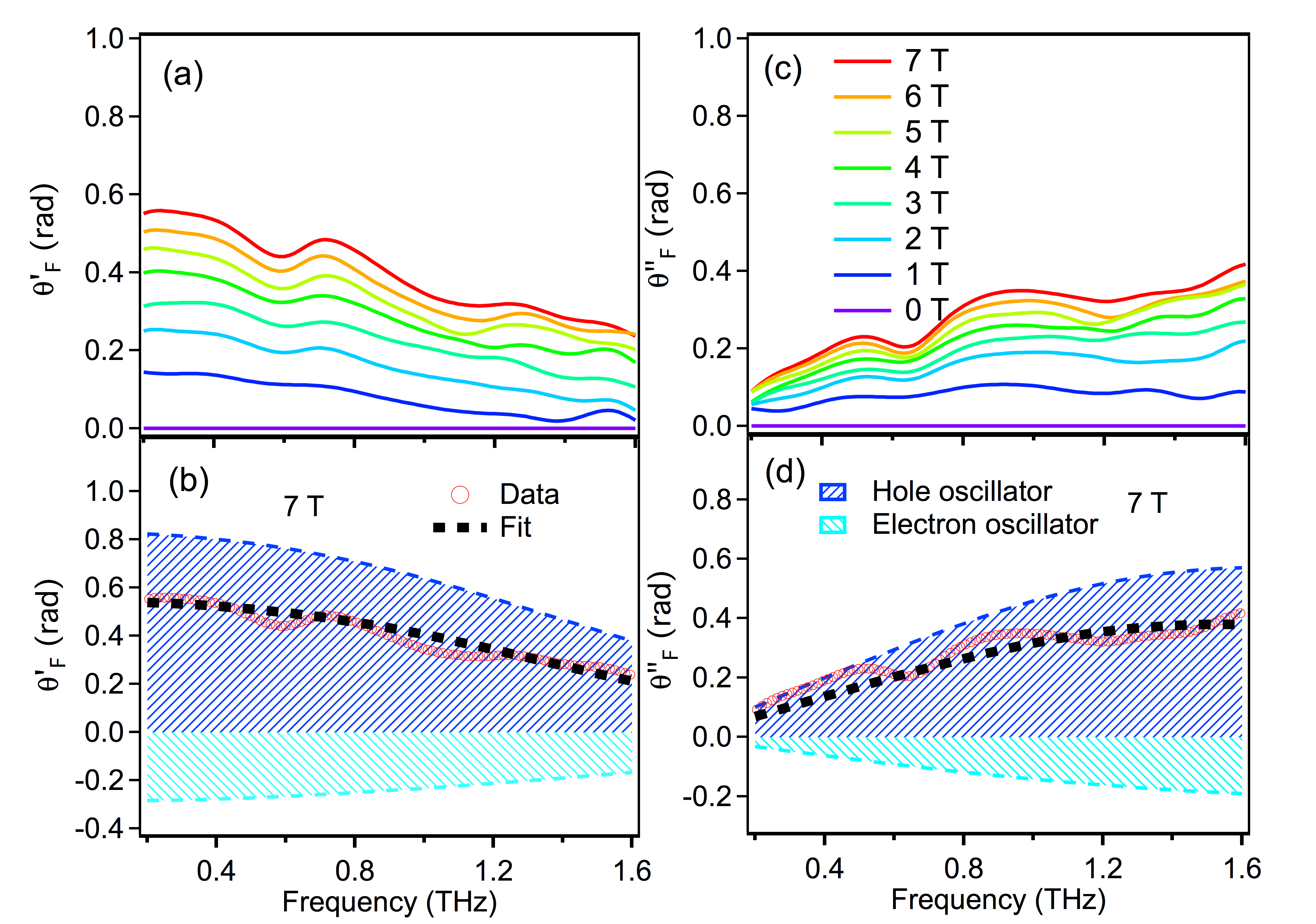}
\caption{(Color online) (a) and (c) Real and imaginary parts of Faraday rotation angle at 6 K. (b) and (d) two Drude fits of real and imaginary parts of Faraday rotation angles at 7 T. }
\end{figure}

To isolate exact CR energies at different fields, we used the Drude model to fit the complex right and left-hand optical conductivities. The expression for the Drude conductivity in magnetic field is

\begin{equation}
\sigma_{\mp}(\omega)=i\epsilon_0\omega  \Big (\sum_{k=1}^{s}{{-\omega_{pk}^2}\over{-\omega^2-i\omega\Gamma_{pk}\pm\omega\omega_c}}-(\epsilon_\infty-1) \Big ).
\label{Drudefit}
\end{equation}

 \noindent In the above expression, $\mp$ correspond to negative and positive frequencies and $\omega_{c}$ is the CR frequency.  When simulating zero-field data, we found one Drude oscillator can well reproduce the complex optical conductivity as shown in Fig. 1(e). However, under magnetic field in the FG, surprisingly no matter how one tunes the adjustable parameters (scattering rate, CR energy, plasma frequency), a single oscillator cannot fit the real and imaginary parts of $\sigma_{cir} $ simultaneously. A similar issue exists in the simulation of Faraday rotation angle.  The complex Faraday rotation angle ($\theta_{F}$ = $-$arctan$[i(T_{r}-T_{l})/(T_{r}+T_{l})]$) can be expressed as a function of right-hand and left-hand optical conductivity respectively as discussed above.  In Fig. 3(a) and 3(c), we show the real  ($\theta^{'}_{F}$) and imaginary parts ($\theta^{''}_{F}$) of the Faraday rotation at different fields.  The real part of the Faraday angle gives the rotation of the light's major axis and the imaginary part determines the ellipticity. With increasing field, the Faraday rotation is enhanced. The positive sign indicates again that the Faraday rotation is dominated by hole carriers.  Similar to the conductivity, we found that fitting with just a single Drude oscillator of hole carriers constrained by the zero-field conductivity cannot simulate both $\theta^{'}_{F}$ and $\theta^{''}_{F}$ well simultaneously.

To reproduce the magneto-optical data, we found that it is necessary to add a second Drude term representing electron carrier to the simulation. To constrain the space of fitting parameters, we fit the complex right/left optical conductivity and complex Faraday rotation angles  varying the scattering rate and cyclotron frequency at each field, but keeping the plasma frequency fixed for all fields.  We show the 7 T results of fitting conductivity in Fig. 2(b) and (c), and the results for Faraday rotation in Fig. 3(b) and 3(d). By carefully tuning parameters, all optical data at 7 T are modeled well.  The plasma frequency of the hole oscillator $\omega_{ph}$/2$\pi$ is $\sim$ 95 THz, and the plasma frequency of the electron oscillator $\omega_{pe}$/2$\pi$ is $\sim$ 75 THz. One can estimate the total plasma frequency $\omega_{p}$/2$\pi$ through $\sqrt{\omega_{ph}^{2}+\omega_{pe}^{2}}$/2$\pi$ as 121 THz, which is close to the plasma frequency from the fittings of the zero-field data (Fig. 1). 

To further investigate the properties of these two oscillators, we plot their CRs as a function of field in Fig. 4(a). One can see that the hole CR dispersion shows strong curvature, which provides evidence for the nontrival dispersion of the hole bands.  The bulk hole carriers of the topological crystalline insulator Pb$_{1-x}$Sn$_{x}$Te can be described as massive Dirac fermions \cite{TCI_Tian13}.  In contrast, the CR of the electrons increases almost linearly with field.

Due to its approximately linear dependence of its cyclotron resonance in field, information about the mass of the electron band can be directly extracted.   The CR frequency of a conventional electron gas is expressed as $\omega_{c}$=$e$B/$m^*$. Through fitting the data as shown in Fig. 4(a), the cyclotron mass of the electrons are found to be approximately 0.2 $m_{0}$, where $m_{0}$ is free electron mass. In contrast, the magnetic field dependence of the hole bands shows that the chemical potential falls in a regime of linear band dispersion.   These are naturally the special inverted bands usually observed in 3D topological insulators.  To lowest approximation these are expected to have a hyperbola-like Dirac dispersion that is gapped at low energy, but has a linear dependence at high energy.  Hence the system can be regarded as a Dirac semimetal with a small band (mass) gap. The CRs of massive Dirac fermions can be expressed as

\begin{equation}
E_{c}=\sqrt{2e\hbar v_F^2B|n|+\Delta^2}-\sqrt{2e\hbar v_F^2B(|n|-1)+\Delta^2}.
\label{CR}
\end{equation}

\noindent The above expression describes intraband LL transitions from the highest occupied $n$th LL to the lowest unoccupied $(n+1)$th LL. Because the Dirac bands are hole doped,  the LL index $n$ related to CRs are negative integers. Here $v_F$ is the Fermi velocity and $\Delta$ is one half of the band gap [Fig. 1(b)]. From Eq. \ref{CR}, if $\left|n\right|$ $\gg$ 1, the CRs will show the quasi-classical behavior and linear field dependence. The strong curvature of hole CRs clearly indicates the system is close to quantum limit and the index $\left|n\right|$ of LLs which take part in the intraband transitions should be small in the high field part. We show the simulations of CRs in Fig. 4(a) plotted alongside. One can see, although the low field CRs are not fitted very well, in high field region, hole CRs can be well described by the intraband transitions from LL($-$3) to LL($-$2). Through the simulations, the Fermi velocity $v_{F}$, the band gap 2$\Delta$ and Fermi energy $E_F$ are estimated to be (2$\pm$0.2)$\times$$10^5$ m/s, 30$\pm$2 meV and $-$(35$\pm$5) meV respectively.  For high enough chemical potentials, we can estimate the effective mass of hole carriers through the formula $E_F = m^*v_F^2$ and $m^*$ is found to be $\sim$ 0.14 $m_0$.

\begin{figure}[t]
\includegraphics[clip,width=3.7in]{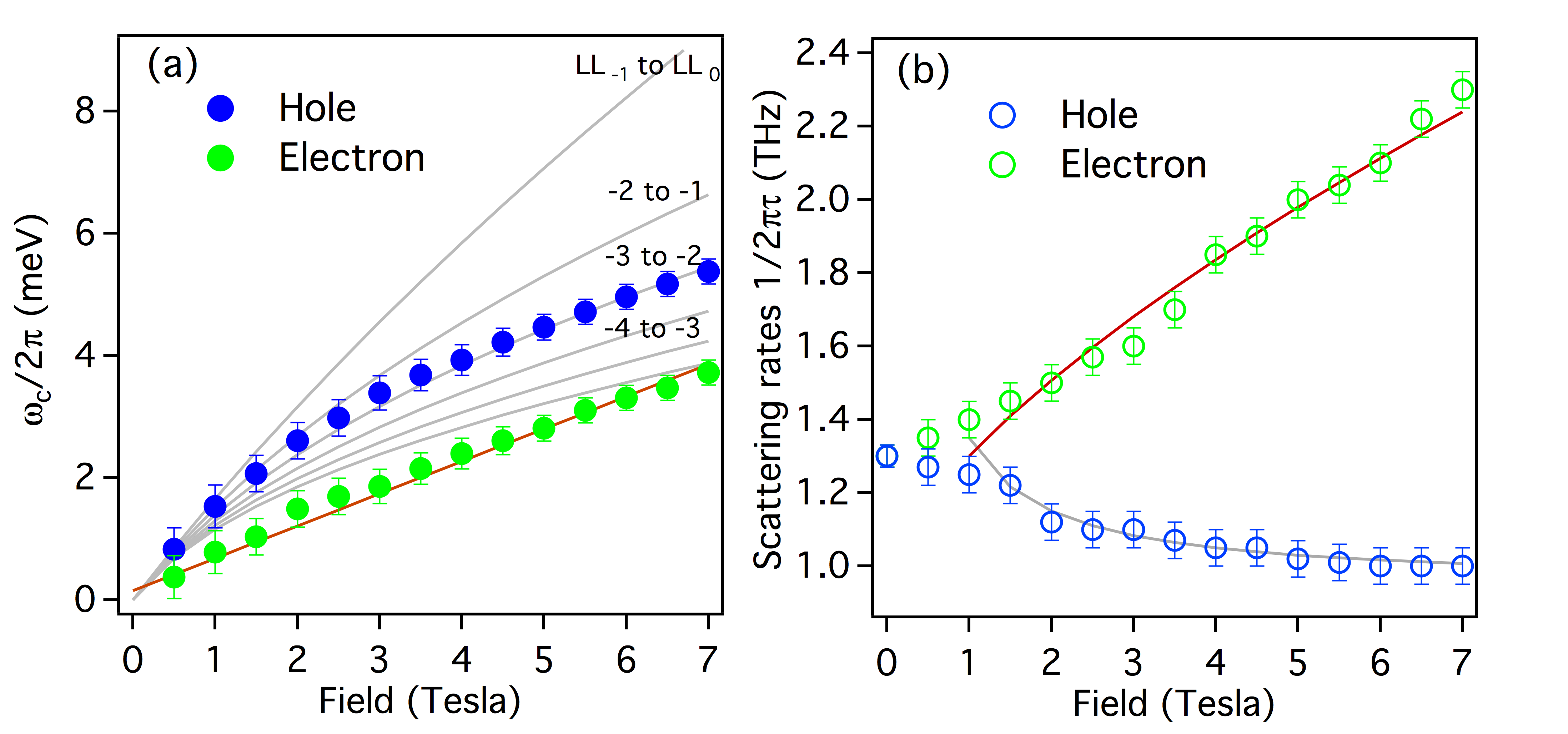}
\caption{(Color online) (a) Cyclotron resonances of holes and electrons as functions of field. The grey (red) curves show the simulations of the cyclotron resonances between different LLs of holes (electrons). (b) Field dependence of scattering rates of free carriers in the Faraday geometry. The red curve is a $a$+$b$$B^{\frac{2}{3}}$ model of the field dependent scattering rates of electrons. Here $a$ and $b$ are positive coefficients. $a$ represents the zero-field impurity scattering strength that is found to be $\sim$ 1 THz. The grey curve is added to guide the eye.}
\end{figure}

The presence of an electron oscillator is notable. Pb$_{0.5}$Sn$_{0.5}$Te is located in the hole doped part of the phase diagram and its free bulk carriers have been reported to be p-type \cite{volobuev2017giant}.  We will also point out that transport measurements have shown a non-linear Hall effect indicating the presence of at least two conductance channels \cite{dziawa2012topological}. In regard to the origin of electron carriers, we may exclude the possibility of TSSs. If the electron Drude oscillator arose from TSSs, with the knowledge of the effective mass and spectral weight, we can estimate that the Fermi energy would have had to be 5 eV above the Dirac points of the TSSs [See supplementary material].  This is clearly inconsistent with other measurements \cite{Okada1496,TCI_optics16}. However, one possibility is that they arising from trivial states, which come from a surface band bending accumulation layer. Such states occur in conventional topological insulators \cite{king2011large}.  Another possibility is that the electron contribution arises from the bulk impurity bands, which have been observed in In-doped SnTe\cite{Cava16_In_TCI,In_TCI_18}. Actually, due to various intrinsic or extrinsic mechanisms, it is very common to observe multi-channel transport in topological semimetals such as Cd$_3$As$_2$\cite{Dirac_Timo_prb}, Na$_3$Bi\cite{Xiong413}, TaAs\cite{TaAs_transport_2015}, ZrTe$_5$\cite{ZrTe5_18}, WTe$_2$\cite{WTe2}, HfTe$_5$\cite{HfTe5}, TaAs$_2$\cite{TaAs2} and NbAs$_2$\cite{NbAs2_16}. In this regard, the coexistence of conventional trivial fermions and Dirac fermions in Pb$_{0.5}$Sn$_{0.5}$Te provides an opportunity to study their different magneto-terahertz responses. At zero field, we find the mobilities of holes and electrons are 1600 and 1100 cm$^{2}$V$^{-1}$s$^{-1}$ respectively. The electron carriers observed in Pb$_{0.5}$Sn$_{0.5}$Te thin films experience low scattering, but are still slightly less mobile than their hole counterparts

Further important information can be derived from the field dependence of the scattering rates. As shown in Fig. 4(b), the scattering rate of holes decreases when switching on the field. In contrast, the scattering rates of electron carriers is an increasing function of field. These distinct field dependences are likely related to the character of their dispersions.  At 6 K, thermal lattice vibrations are largely suppressed and the scattering of the system is mainly from disorder potentials. It has been predicted in graphene (which hosts two-dimensional massless Dirac fermions), that for high enough fields the widths of LL broadening $\Gamma$ may have $\frac{1}{B}$ dependent charged impurity scattering due to the special features of the LLs \cite{LL_scattering10}. In our case, hole Dirac bands will form one-dimensional LL structures which will be dispersive in $k_{z}$ direction, however the principle magneto-optical terahertz response will be from the LLs at $k_{z}$ = 0 due to the singularities in the joint density of states there, making the 2D expression relevant.  In accord with this general dependence we indeed find that the scattering rate of the hole band is a decreasing function of field, which supports its unconventional nature.  This decreasing dependence with field itself can be an indicator of Dirac fermions but has been rarely observed in Dirac systems because other scattering mechanisms, such as CR-phonon coupling \cite{LiangWu_phonon_2015} or strong electronic correlation (when the charge density is low \cite{Graphene_scattering_2011}), may intervene and mask the intrinsic field dependence. For example, in a well-known topological insulator Bi$_{2}$Se$_{3}$, the CR of surface Dirac fermions has a strong coupling with phonons when the CR energy matches the surface $\beta$ phonon around 0.8 THz. The scattering rates of its Drude oscillator is modified to be an increasing function of field \cite{LiangWu_phonon_2015}. The observation of the decreasing dependence of hole scattering rates on field in Pb$_{0.5}$Sn$_{0.5}$Te thin film indicates this system is spectroscopically clean due to the absence of clear signature of optical phonon below 2 THz and can be used as a reference for studying the general behavior of Dirac and Weyl systems. In contrast, the electron carriers of the trivial bands do not have associated Berry curvature and other features of the Dirac fermions and undergo larger scattering. Their scattering rates show quasi-linear field dependence. As shown in Fig. 4(b) the field dependence can also be approximately described by the impurity scattering mechanism in conventional 3D electron gas, which is predicted to exhibit a $a + b B^{\frac{2}{3}}$ field dependence \cite{scatt_3D}.   Our data is consistent with this expression although there is large error bar on the precise exponent.

Finally we wish to make some remarks on the general relevance of our work to current transport investigations on 3D topological semimetals.  Our work provides a new way to isolate signatures of bulk states in Dirac and Weyl semimetals.  Despite the fact that our system's band structure is gapped it is clear that the chemical potential is in a part of the spectrum where the dispersion is linear and as such the terahertz response of this band should be indistinguishable from a gapless Dirac system.  However, as compared to topological insulators such as Bi$_{2}$Se$_{3}$, such 3D topological semimetals usually have more complicated electronic structures. Multiple bands may cross the Fermi level, making the Fermi surfaces potentially complex \cite{TaAs_LDA15,TaAs_LDA15_2}.  It has been reported that multiple types of carriers coexist in some 3D topological semimetals \cite{Dirac_Timo_prb,Xiong413,TaAs_transport_2015}. The ultrahigh mobilities of carriers in these topological semimetals give rich behavior of the magnetoresistivity \cite{ZrTe5_18,WTe2,HfTe5,TaAs2,NbAs2_16}, but in some systems that are known to be topologically trivial, carriers can also have large mobilities and show magnetorestivity responses similar to topological semimetals \cite{a-WP2}.   The confusing experimental situation calls for more comprehensive probes in the field of topological materials and non-trivial band geometry. Magneto-terahertz spectroscopy with its field and frequency dependence provides enough information to characterize mobilities and densities of multiple transport channels simultaneously. Even more importantly, cyclotron resonance and field-dependent scattering rates, which directly store information about non-trivial band geometry and Berry curvature, can be extracted from circular-basis magneto-terahertz conductivity for each transport channel. Moreover, the non-contact nature of terahertz measurement can avoid many of the artifacts that may exist in conventional dc transport measurements such as extrinsic current jetting \cite{NPA18}. These advantages provide an opportunities to solve puzzles in the field of topological semimetals such as the origins of negative magnetoresistances and intrinsic signatures of the chiral anomaly \cite{NPA18,Dirac_Timo_prb}.
 
We would like to thank Dipanjan Chaudhuri and Y. C. Wang for helpful discussions.  Experiments at JHU were supported by the Army Research Office Grant W911NF-15-1-0560. This research was partially supported by a Laboratory University Collaboration Initiative award provided by the Basic Research Office in the Office of the Under Secretary of Defense for Research and Engineering.   

 \bibliography{Quadratic}
 
 \newpage

\setcounter{figure}{0}
\setcounter{equation}{0}
\setcounter{section}{0}
\begin{widetext}

\title{Supplemental Material: Magneto-Terahertz Response and Faraday Rotation from Massive Dirac Fermions in Topological Crystalline Insulator Pb$_{0.5}$Sn$_{0.5}$Te }

\maketitle
\title{Supplemental Material: Magneto-Terahertz Response and Faraday Rotation from Massive Dirac Fermions in Topological Crystalline Insulator Pb$_{0.5}$Sn$_{0.5}$Te }

\section{Supplementary Note 1:  Thin-film growth and terahertz spectroscopy }

(111)-oriented Pb$_{0.5}$Sn$_{0.5}$Te thin films were grown epitaxially on (100) GaAs single crystal substrates by solid source molecular beam epitaxy (MBE), using pure elements as sources, to a range of different thicknesses.  Complex values of the transmission matrix $T_{xx}$ and $T_{xy}$ in THz range were measured by a home-built TDTS spectrometer in a closed-cycle 7 T superconducting magnet.   The complex conductivity of the films can be directly extracted in the thin-film limit with the expression: $T(\omega) = \frac{1+n}{1+n+Z_{0}d\sigma(\omega)}\mathrm{exp}[\frac{i\omega}{c}(n-1)\Delta L]$. Here $T(\omega)$ is the transmission of a particular eigenpolarization as referenced to GaAs substrate, $\sigma$($\omega$) is the complex optical conductivity in the corresponding basis, $d$ is the film thickness, and $n$ is the index of refraction of the substrate and $Z_{0}$ is the vacuum impedance.  A fast rotating polarizer (FRP) setup was used to modulate the polarization of terahertz pulses being transmitted through the sample, allowing the polarization of the pulse to be determined to high accuracy in a single measurement. With the knowledge of the polarization state of the transmitted THz pulse, one can calculate the complex Faraday rotation and the optical conductivity in the circular basis as described below.  

\begin{figure}[b]
\includegraphics[clip,width=3.6in]{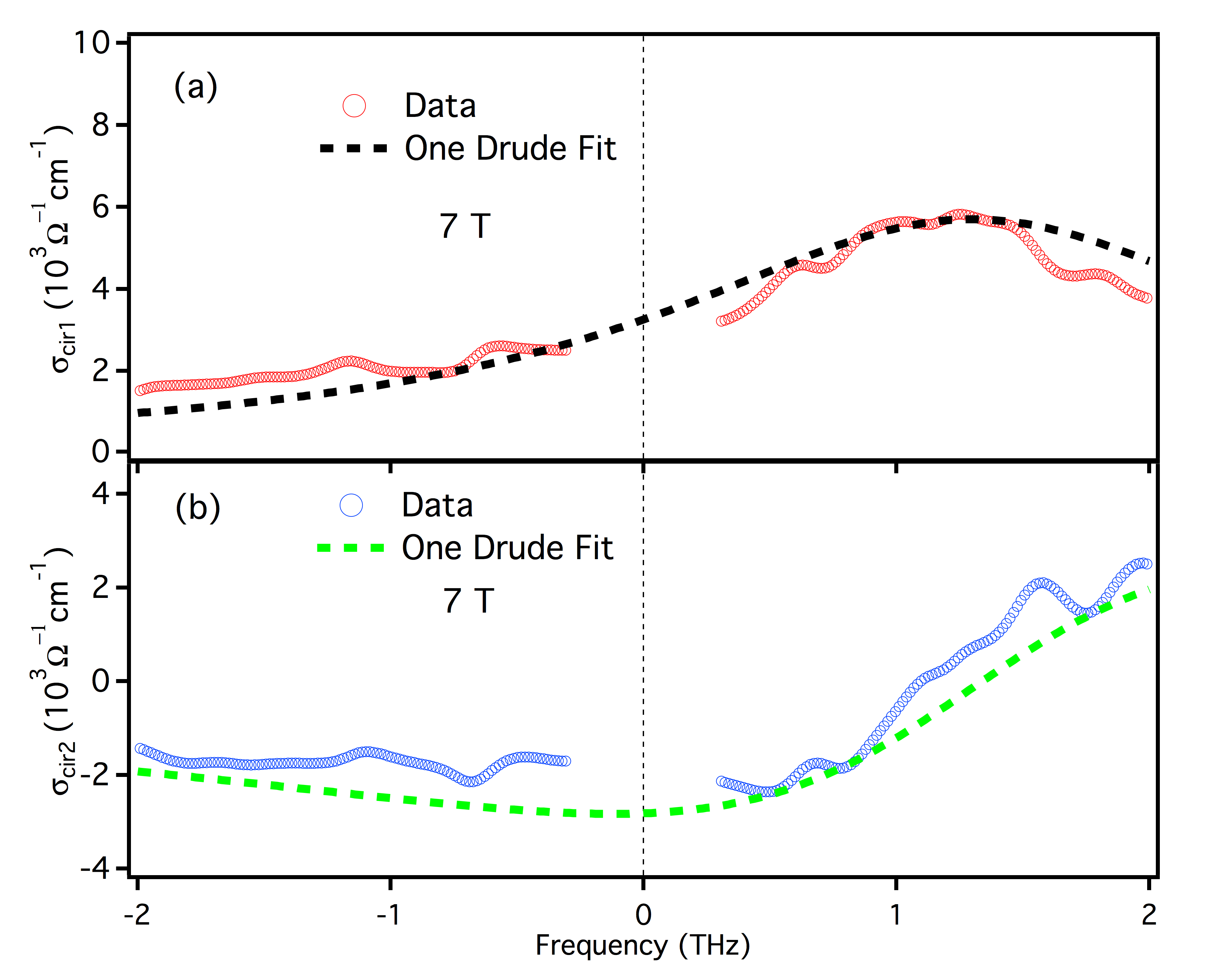}
\caption{(Color online) One Drude fits for 7 T real part (a) and imginary part (b) optical conductivity in circular basis.  }
\label{xxx}
\end{figure}

\begin{figure}[t]
\includegraphics[clip,width=3.6in]{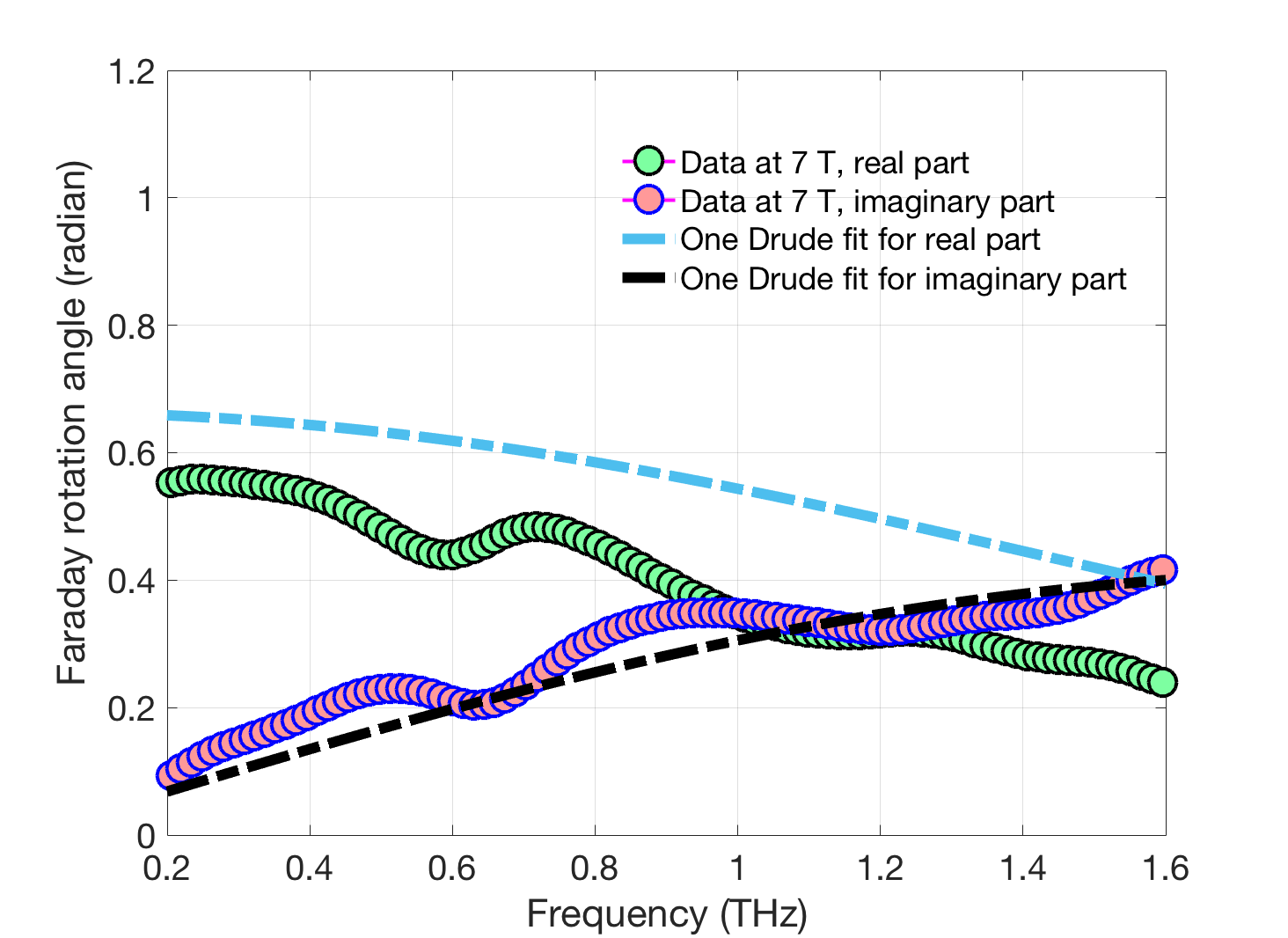}
\caption{(Color online) One Drude fits for 7 T real part (a) and imginary part (b) Faraday rotation angle.  }
\label{xxx}
\end{figure}

\begin{figure}[t]
\includegraphics[clip,width=3.6in]{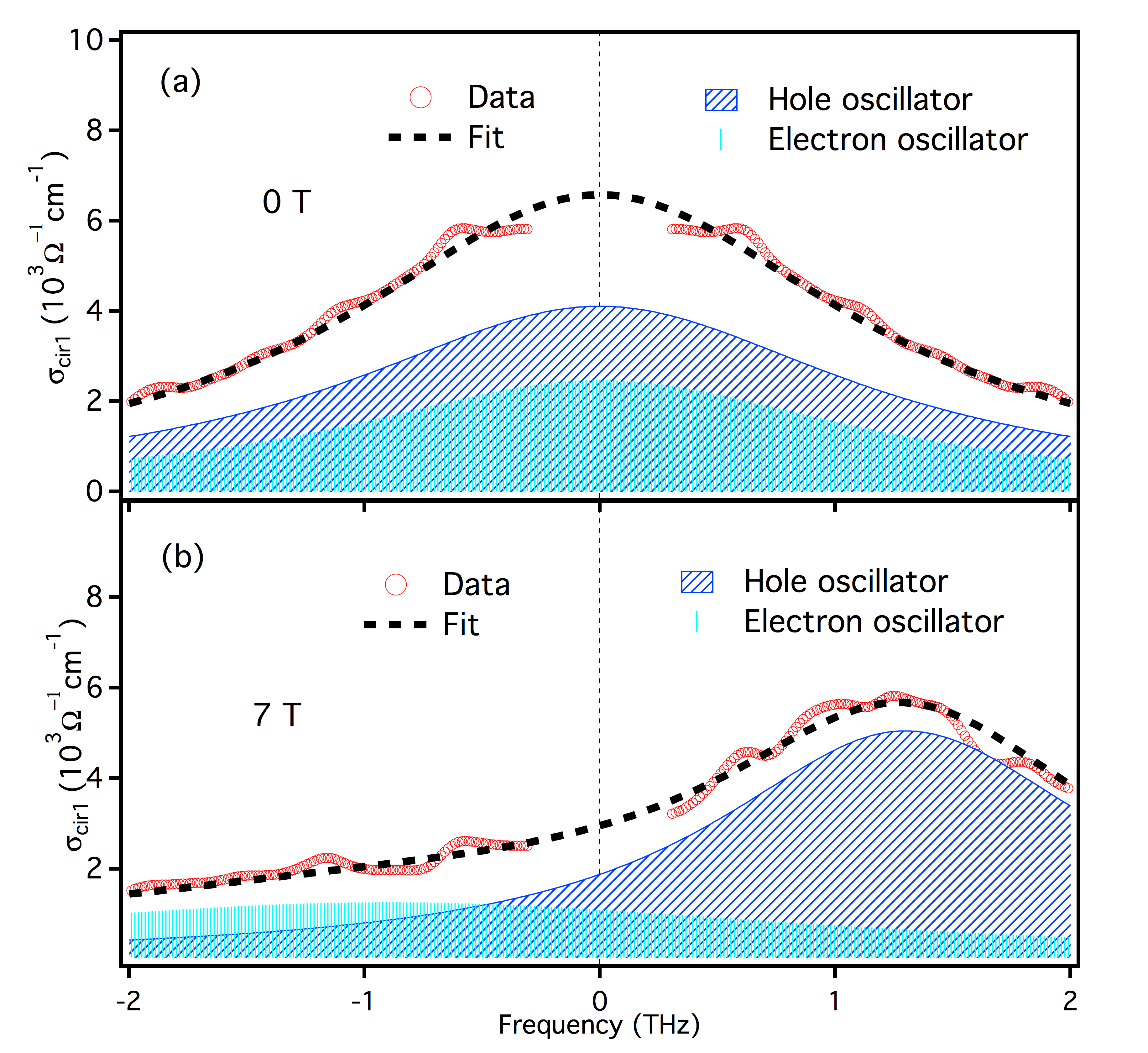}
\caption{(Color online) Two Drude fits for 0 T (a) and 7 T (b) real part optical conductivity in circular basis.  }
\label{xxx}
\end{figure}

\section{Supplementary Note 2:  Calculations of optical conductivity in circular polarization basis }

In the Faraday geometry, the magnetic field is perpendicular to the sample surface and the linear polarization basis is not the eigenbasis of the transmitted terahertz beam.   If a $C_4$ or $C_3$ symmetry exists, then the eigenpolarization basis will be circular.  Then to calculate the optical conductivity, we need to transfer the measured transmission $T_{xx}$ and $T_{xy}$ to left-hand transmission $T_l$ and right-hand transmission $T_r$ coefficients through the expression

\begin{equation}
\hat{T}_{cir}={
\left[ \begin{array}{ccc}
T_r  & 0\\
0  & T_l\\

\end{array} 
\right ]}=
{
\left[ \begin{array}{ccc}
T_{xx}+iT_{xy} & 0\\
0 & T_{xx}-iT_{xy}\\

\end{array}
\right ].}
\end{equation}

\noindent The complex conductivity of the film in circular polarization basis can be directly extracted in the thin-film limit with the expression: $T(\omega) = \frac{1+n}{1+n+Z_{0}d\sigma(\omega)}$exp$[\frac{i\omega}{c}(n-1)\Delta L]$. Here $T(\omega)$ is the left or right hand transmission as referenced to GaAs substrate. $\sigma$($\omega$) is the left or right hand complex optical conductivity. $d$ is the film thickness, and $n$ is the refraction index of substrate. $\Delta L$ is the small thickness difference between samples and reference substrates, and $Z_{0}$ is the vacuum impedance, which is approximately 377 $\Omega$.

\section{Supplementary Note 3:  Fits of optical conductivity and Faraday rotation }

When simulating zero-field data, we found one Drude oscillator can well reproduce the complex optical conductivity as shown in the main text's Figure 1. Considering that the CR exhibited prominently in the right-hand conductivity comes from hole carriers and the sample is located at the hole-doping region in the phase diagram, it is natural to try to fit with a single oscillator, which contains the entire spectral weight ($\sim$ 124 THz) of the zero field simulation to reproduce the optical conductivity in both channels. In supplementary Figure 1 we show one Drude fit for 7 T real and imaginary conductivity plotted with R and L handed polarizations as positive and negative frequencies respectively, with fixing the spectral weight about 124 THz. One can see that, by using one Drude oscillator, no matter how we tune the scattering rate, the data at 7 T cannot be fit well.

 A similar issue exists in the simulation of Faraday rotation angle. The complex Faraday rotation angle from a simple expression: $\theta_{F}$ = $-$arctan$[i(T_{r}-T_{l})/(T_{r}+T_{l})]$. $T_{r}$ and $T_{l}$ are right hand and left hand transmissions respectively and can be expressed as a function of $\sigma_{r}$($\omega$) and $\sigma_{l}$($\omega$) through the formula: $T(\omega) = \frac{1+n}{1+n+Z_{0}d\sigma(\omega)}exp[\frac{i\omega}{c}(n-1)\Delta L]$.  Similar to the conductivity, as shown in supplementary Figure 2, we found that fitting with just a single Drude oscillator of hole carriers constrained by the zero-field conductivity cannot reproduce both $\theta^{'}_{F}$ and $\theta^{''}_{F}$ simultaneously.

As shown in the main text Figure 2 and 3, we need two Drude terms (one of which represents hole carriers and the other represents electron carriers) to reproduce the Faraday rotation angle and optical conductivity well. To see clearly the motion of these two Drude terms under magnet field, we plot two Drude fits for 0 T and 7 T optical conductivity together. One can see the hole Drude oscillator moves along the positive frequency axis and the electron oscillator moves along the negative frequency axis. Furthermore, with increasing field, the width of hole oscillator decreases and the width of electron oscillator increases.

\section{Supplementary Note 4:  Landau Level simulations }

The Landau level energies of 3D massive Dirac system can be expressed in the below form

\begin{equation}
\begin{aligned}
E_{n}(k_{z})=\pm\delta_{n,0}\sqrt{(\hbar vk_z)^2+\Delta^2}\\
+\textrm{sgn}(n)\sqrt{2e\hbar v^2B|n|+(\hbar vk_z)^2+\Delta^2}.
\label{chik}
\end{aligned}
\end{equation}

\noindent In the above expression, $n$ is Landau level index. $\Delta$ is one half the band gap and $v$ is the band velocity at large energies. Positive/negative sign of $n$ represents Landau level of conduction/valence bands. In principal, the intraband Laudau level (LL) transition at $k_{z}$ = 0 will contribute most of the spectral weight to the CR due to the singularities in the joint density of states. The formula of CR energy can then be expressed as
\begin{equation}
E_{c}=\sqrt{2e\hbar v^2B|n|+\Delta^2}-\sqrt{2e\hbar v^2B(|n|-1)+\Delta^2}.
\label{Ec}
\end{equation}

\noindent As we show in the main text, the hole carriers come from 3D Dirac bands. The CRs show strong curvature as a function of field, which indicates the Fermi level should be not be too high. In a previous study  of Pb$_{x}$Sn$_{1-x}$Te with a doping similar to our sample, $\Delta$ is reported to be 15 meV. Then to simulate CRs of holes, we fix $\Delta$ around 15 meV. The detailed simulation is shown in the main text Figure 3. We show the simulations of Landau levels in supplementary Figure 4. The shadow area indicates the possible positions of the Fermi level. 
\begin{figure}[b]
\includegraphics[clip,width=3.6in]{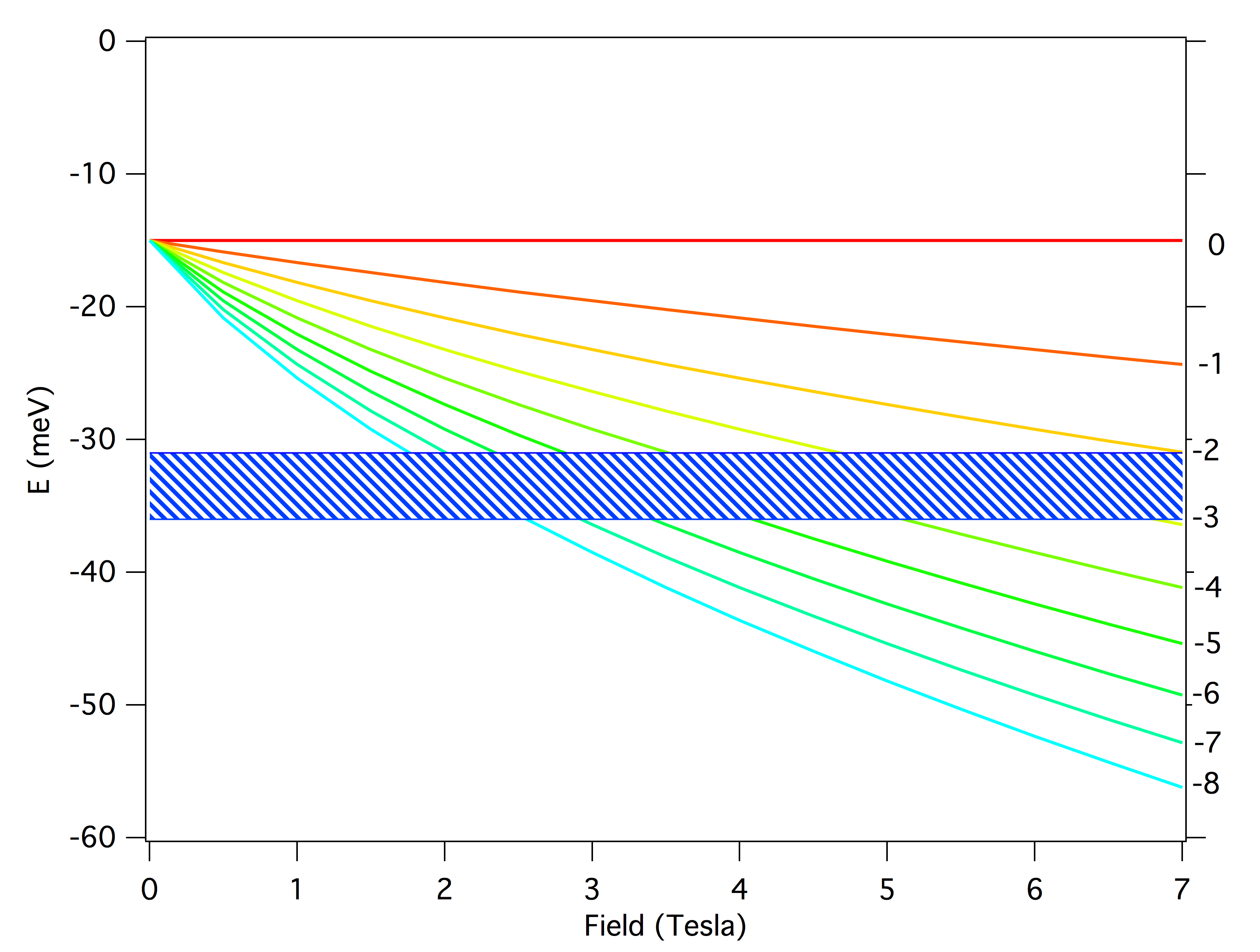}
\caption{(Color online) Landau level simulations with $v_F$ = 2$\times$$10^5$ m/s and $\Delta$ = 15 meV. The shadow area indicates the possible positions of Fermi level which can make intraband transibiton between LL(-3) to LL(-2) possible in high field region.}
\label{xxx}
\end{figure}

\section{Supplementary Note 5:  Optical spectral weight analysis for hole bands}

 When we simulate R/L-hand optical conductivity, the plasma frequency of these hole carriers is found to be $\omega_{ph}/2\pi \sim 95$ THz, where $\omega_{ph}=\sqrt{ne^{2}/m^{*}\epsilon_{0}}$. The classic Luttinger theorem correlates the charge density $n$ with the Fermi vector $k_{F}$ via $n = 4k_{F}^3/3\pi$. Here, the spin degeneracy 2 and valley degeneracy 4 are taken into account. The effective mass is estimated to be 0.14 $m_0$ through simulate LLs and CRs. The charge density $n$ for holes is $\sim$ 1.5 $\times$ $10^{19}$ cm$^{-3}$. Then the Fermi vector $k_F$ is estimated to be $\sim$ 3 $\times$ $10^8$ m$^{-1}$. In a massive 3D Dirac system, the Fermi energy $E_F$ can be calculated through the expression $E_{F}=-\sqrt{\hbar^{2}v_F^2k_{F}^2+\Delta^{2}}$.  $E_F$ is found to be $\sim$ $-$40 meV, which is very close to the value estimated ($-$35 meV)by LL and CR simulations. The small discrepancy may arise from anisotropies of the bulk Fermi surfaces.

 \section{Supplementary Note 6: Optical spectral weight analysis for electron bands }

 The plasma frequency of the electron Drude term was extracted to be $\sim$ 75 THz. Considering the bulk band gap is around 30 meV, topological surface states (TSSs) probably make some contribution to the observed spectral weight as the Dirac points of the surface bands are located in bulk valence bands slightly. Here, we demonstrate the main spectral weight of the electron Drude term should not arise from the TSSs. From fitting the electron CRs, the effective mass is estimated to be $\sim$ 0.2 $m_0$. We can use $\omega_{pn}=\sqrt{ne^{2}/m^{*}\epsilon_{0}}$ to estimate the 2D charge density of electrons.  We find it to be $\sim$ 4.5 $\times$ 10$^{14}$ cm$^{-2}$. TSSs have an approximately linear energy dispersion and the 2D charge density can be calculated by $n_{2D}$ = $k_F^2$/$\pi$. Here, we already account for the fact that the (111) surface Brillouin zone has four surface Dirac bands and assume $k_F$ = 0 if surface Fermi level cross Dirac points of TSSs. If all the observed spectral weight of the electron Drude came from the TSSs, the Fermi wave vector $k_F$ would be $\sim$ 3.7 $\times$ $10^9$ m$^{-1}$. Finally, we could use the expression $E_F$ = $\hbar^2 k_F^2$/$m^*$ to estimate $E_F$ of TSSs about $\sim$ $-$5 eV. This means the Dirac points of TSSs should be 5 eV below the bulk valence band edge, which is evidently not possible. These calculations also indicate in this thick sample (325nm), bulk states should be dominant. Even if the TSSs can make a contribution to a part of the optical spectral weight, as compared to the bulk spectral weight, the surface spectral weights should be negligible. To see this, let us assume the Fermi energy of TSSs is $-$500 meV (This value is also too big for the reasonable Fermi level of TSSs). With the same effective mass of electron carriers, the spectral weights of TSSs will be one tenth of the spectral weights of electron Drude. The TSS spectral weights must be less than 5 \% of the total observed Drude spectral weight.

\end{widetext}

\end{document}